\documentclass[%
 reprint,
 amsmath,amssymb,
 aps,
nofootinbib,
]{revtex4-2}
\usepackage{graphicx}
\usepackage{epstopdf}
\usepackage{color}
\usepackage{multirow}
\usepackage{subeqnarray}
\usepackage{setspace}
\usepackage{palatino}
\usepackage{dcolumn}
\usepackage{bm}
\usepackage[utf8]{inputenc}
\usepackage[T1]{fontenc}
\usepackage{mathptmx}
\usepackage{amsmath,amsfonts,amscd,amssymb,amsthm}
\usepackage{commath}

\usepackage{algorithm,algcompatible,algorithmicx}
\usepackage{algpseudocode}

\algnewcommand\algorithmicinput{\textbf{Input:}}
\algnewcommand\INPUT{\item[\algorithmicinput]}
\algnewcommand\algorithmicoutput{\textbf{Output:}}
\algnewcommand\OUTPUT{\item[\algorithmicoutput]}
\algnewcommand\algorithmicoptional{\textbf{Optional:}}
\algnewcommand\OPTIONAL{\item[\algorithmicoptional]}
\usepackage{overpic}
\usepackage{cancel}
\usepackage{rotating}
\usepackage{url}
\usepackage{caption}
\usepackage{subcaption}
\usepackage{mathtools}

\DeclareMathOperator*{\argmin}{arg\,min}
\usepackage{setspace}
\usepackage[bottom,flushmargin,hang,multiple,symbol]{footmisc}
\newcommand\blfootnote[1]{%
  \begingroup
  \renewcommand\thefootnote{}\footnote{#1}%
  \addtocounter{footnote}{-1}%
  \endgroup
}

\begin{document} 

\title{Permanent magnet optimization for stellarators as sparse regression}

\author{Alan A. Kaptanoglu}\thanks{Corresponding author (akaptano@umd.edu).}
 \affiliation{Institute for Research in
Electronics and Applied Physics,\\ University of Maryland, College Park, MD, 20742, USA\looseness=-1} 
\author{Tony Qian}
\affiliation{Princeton Plasma Physics Laboratory,\\ 100 Stellarator Road, Princeton, NJ, 08540, USA\looseness=-1}
\author{Florian Wechsung}
 \affiliation{Courant Institute of Mathematical Sciences,\\ New York University, New York, NY, 10012, USA\looseness=-1} 
\author{Matt Landreman}
 \affiliation{Institute for Research in
Electronics and Applied Physics,\\ University of Maryland, College Park, MD, 20742, USA \looseness=-1} 

 \begin{abstract}
 A common scientific inverse problem is the placement of magnets that produce a desired magnetic field inside a prescribed volume. This is a key component of stellarator design, and recently permanent magnets have been proposed as a potentially useful tool for magnetic field shaping. 
 Here, we take a closer look at possible objective functions for permanent magnet optimization, reformulate the problem as sparse regression, and propose an algorithm that can efficiently solve many convex and nonconvex variants. The algorithm generates sparse solutions that are independent of the initial guess, explicitly enforces maximum strengths for the permanent magnets, and accurately produces the desired magnetic field. The algorithm is flexible, and our implementation is open-source and computationally fast. We conclude with two new permanent magnet configurations for the NCSX and MUSE stellarators. 
 Our methodology can be additionally applied for effectively solving permanent magnet optimizations in other scientific fields, as well as for solving quite general high-dimensional, constrained, sparse regression problems, even if a binary solution is required. 
 \end{abstract}

 \maketitle

\section{Introduction}\label{sec:intro}
Magnet design is required in a large number of scientific domains, but it is a fundamentally ill-posed problem because many different magnet designs can produce an identical target magnetic field via the Biot-Savart law. An extreme case for magnetic design comes from plasma experiments that are investigating nuclear fusion; these experiments often require very strong and complex magnetic coils. One class of plasma experiments, stellarators, particularly relies on sophisticated coil design algorithms in order to produce ideal magnetic fields for confining plasma~\cite{grieger1992physics,imbert2019introduction}. These three-dimensional magnetic fields must be carefully shaped in order to provide high-quality confinement of charged particle trajectories and many other physics objectives. Stellarator optimization is usually divided into two stages. The first is a configuration optimization using fixed-boundary magnetohydrodynamic (MHD) equilibrium codes to obtain MHD equilibria with desirable physics properties~\cite{drevlak2018optimisation,lazerson2020stellopt,landreman2021simsopt}. 

After obtaining the optimal magnetic field in this first stage, coils must be designed to produce these fields, subject to a number of engineering constraints such as a minimum coil-to-coil distance, maximum forces on the coils~\cite{robin2022minimization}, maximum curvature on the coils, and many other requirements~\cite{zhu2017new}. The result is that stellarator coils are often very complex 3D shapes, raising the cost and difficulty of manufacturing. The primary cost of the W-7X stellarator program was the manufacture of these complex coils with very tight engineering tolerances~\cite{erckmann1997w7}. The NCSX stellarator was never finished in large part because of similar fabrication and assembly obstacles~\cite{strykowsky2009engineering}. 

\subsection{A role for permanent magnets}\label{sec:permanent_magnets_intro}
Until recently, the highly-shaped and precise nature of stellarator equilibria implied the necessity of these very complex coils. One way to circumvent this requirement, first proposed in Helander et al.~\cite{helander2020stellarators}, is to simplify stellarator coil designs by surrounding a stellarator with a manifold of permanent magnets that can provide significant portions of the magnetic field. These permanent magnets cannot be used to generate a net toroidal flux, so traditional magnetic coils are still required. Instead, the permanent magnets allow for significant reductions in the coil complexity and cost. Moreover, permanent magnets operate without power supplies, require minimal cooling, ameliorate magnetic ripple due to discrete coils, and facilitate improved diagnostic access to the plasma chamber. However, there are also some potential disadvantages, including the inability to turn off the field, the possibility of demagnetization, and an upper limit on the achievable field strength; material science advances~\cite{wang2020environment} for permanent magnets could significantly address the latter two disadvantages. Despite these potential setbacks, the low cost and simple manufacture of permanent magnet stellarators are tantalizing, especially for university-lab-scale experiments.
There is already a promising experimental effort to produce a very cheap but practical permanent magnet stellarator~\cite{qian2021stellarator}. In the present work, we will show that permanent magnet stellarators have an additional advantage. Unlike filamentary coil optimization, the permanent magnet optimization problem can be expressed as sparse regression. Thanks to the prolific scientific interest in variations of sparse regression, permanent magnet optimization can subsequently be relatively well understood and rapidly solved.
Moreover, traditional stellarator coil optimization using a winding surface~\cite{landreman2017improved}, even with complicated requirements on coil-coil forces~\cite{robin2022minimization}, can also be formulated as sparse regression. 

\subsection{Motivation}\label{sec:motivation}
Simplifying coils by utilizing permanent magnets comes with its own challenges.
Sophisticated algorithms are still required to find high-quality configurations of permanent magnets. Optimization for determining the optimal placement of permanent magnets (subject to minimizing the cost of the magnets and various engineering constraints) is in a somewhat early stage. There are currently several different formulations and associated algorithms for addressing the permanent magnet optimization problem~\cite{zhu2020topology,zhu2020designing,landreman2021calculation,xu2021design,lu2022development,qian2022simpler}, but the relationships between them are often unclear. Some of the optimization problems are multi-stage or use discrete optimization, and the ``best'' set of loss terms is an open question. These nonconvex problems seem to exhibit many local minima and high sensitivity to the initial conditions. Often additional post-processing optimization steps are taken to further improve the initial optimization solutions. Despite these methods relying on some heuristics and post-processing, they often produce very high-quality solutions. Similarly, genetic algorithms have been used to generate permanent magnet configurations in the magnetic resonance imaging (MRI) community~\cite{cooley2017design,ren2018design}. Here too, there have been few conclusions about optimality, and the algorithms are not always extensible when future work necessitates higher-dimensional problems or additional constraints. 

If the permanent magnet optimization could be better understood and more efficiently solved in the stellarator community, ``off-the-shelf'' permanent magnet stellarators could be rapidly designed for cheap university-lab-scale experiments with very simple toroidal field (TF) coils. If such stellarators can be constructed and widely distributed, stellarator expertise could be rapidly cultivated and the parameter space of quasi-symmetric and quasi-isodynamic stellarators could be rapidly explored. This is especially important, as multi-stage stellarator optimization has recently generated highly quasi-symmetric configurations~\cite{landreman2022magnetic,wechsung2022precise} and near-axis expansions~\cite{jorge2020construction,jorge2020near,jorge2020use,landreman2020magnetic,rodriguez2021solving} have facilitated efficient numerical explorations and new discoveries within a very large parameter space of quasi-symmetric stellarator configurations. Outside the stellarator optimization community, improvements in understanding and algorithms for permanent magnet optimization problems would have significant repercussions for better permanent magnet configurations in MRI, automobile design, and other industrial uses.

\subsection{Contributions of the present work}\label{sec:contributions}
Towards rapid design and distribution of cheap permanent magnet stellarators, this work provides a foundation for understanding the permanent magnet optimization problem by showing it can be reformulated as sparse regression. Then, we provide a computationally efficient, easy-to-use, generally applicable, and open-source code for computing permanent magnet configurations for stellarators. The algorithm produces sparse solutions that are independent of the initial guess, explicitly enforces hard constraints on the dipole moment magnitudes, and illustrates that continuous optimization is sufficient for generating high-performance permanent magnet stellarators. As far as we are aware, it joins a very small and recent list of sparse regression algorithms that can effectively solve problems with $\mathcal{O}(10^6)$ optimization variables, and may be the first algorithm capable of handling a similar number of convex constraints and producing binary solutions. We can regularly solve permanent magnet optimizations that depend on dense matrices of billions of elements.

In particular, we provide a new relax-and-split method for general permanent magnet optimization in the open-source SIMSOPT code~\cite{landreman2021simsopt}. Unlike previous work, the entirety of the permanent magnet pipeline, i.e. the geometry, optimization tools, post-processing, etc., is contained in this single open-source tool. The effectiveness of this new algorithm is demonstrated by finding and illustrating two high-performance permanent magnet stellarator configurations. All of the present work's methodology and results can be found in the SIMSOPT code.

\subsection{Permanent magnet optimization as sparse regression}\label{sec:relationship_with_sparse_regression}

The sparsity-promoting optimization problems that appear in this work are the foundation of the field of sparse regression, which encompasses sparse system identification~\cite{brunton2016pnas,champion2020unified,kaheman2020sindy}, compressed sensing~\cite{eldar2012compressed}, and many other tasks in signal and image processing~\cite{bruckstein2009sparse}. This fundamental relationship with optimization problems occurring in sparse regression means that efficient and well-understood algorithms can be immediately brought to bear on the permanent magnet problem. 

There are two primary differences between the optimization problem that we solve in this paper and the many corresponding sparse regression  problems in other scientific fields. First, the latter typically has $\mathcal{O}(10^2)$ or fewer unknowns, while realistic permanent magnet coil optimizations can easily have $\mathcal{O}(10^5)$ unknowns (with a corresponding $\mathcal{O}(10^5)$ constraints). It follows that, unlike most sparse regression applications, permanent magnet optimization requires algorithms that scale well with the number of unknowns and the number of constraints. A notable exception is that sparse regression for image and signal processing is often very high-dimensional, and in this case approaches to this problem typically either (1) convexify the problem, e.g. with the $l_1$ norm, or (2) use a ``greedy'' algorithm that iteratively and rapidly solves the problem, albeit often with weak guarantees on optimality~\cite{bruckstein2009sparse}. Both of these approaches can struggle to solve binary problems, e.g. the desired solution for permanent magnet optimization is not just a sparse set of magnets but one in which the magnets are all either \textit{maximum-strength} or \textit{exactly} zero.

\begin{figure*}
    \centering
    \begin{overpic}[width=0.85\linewidth]{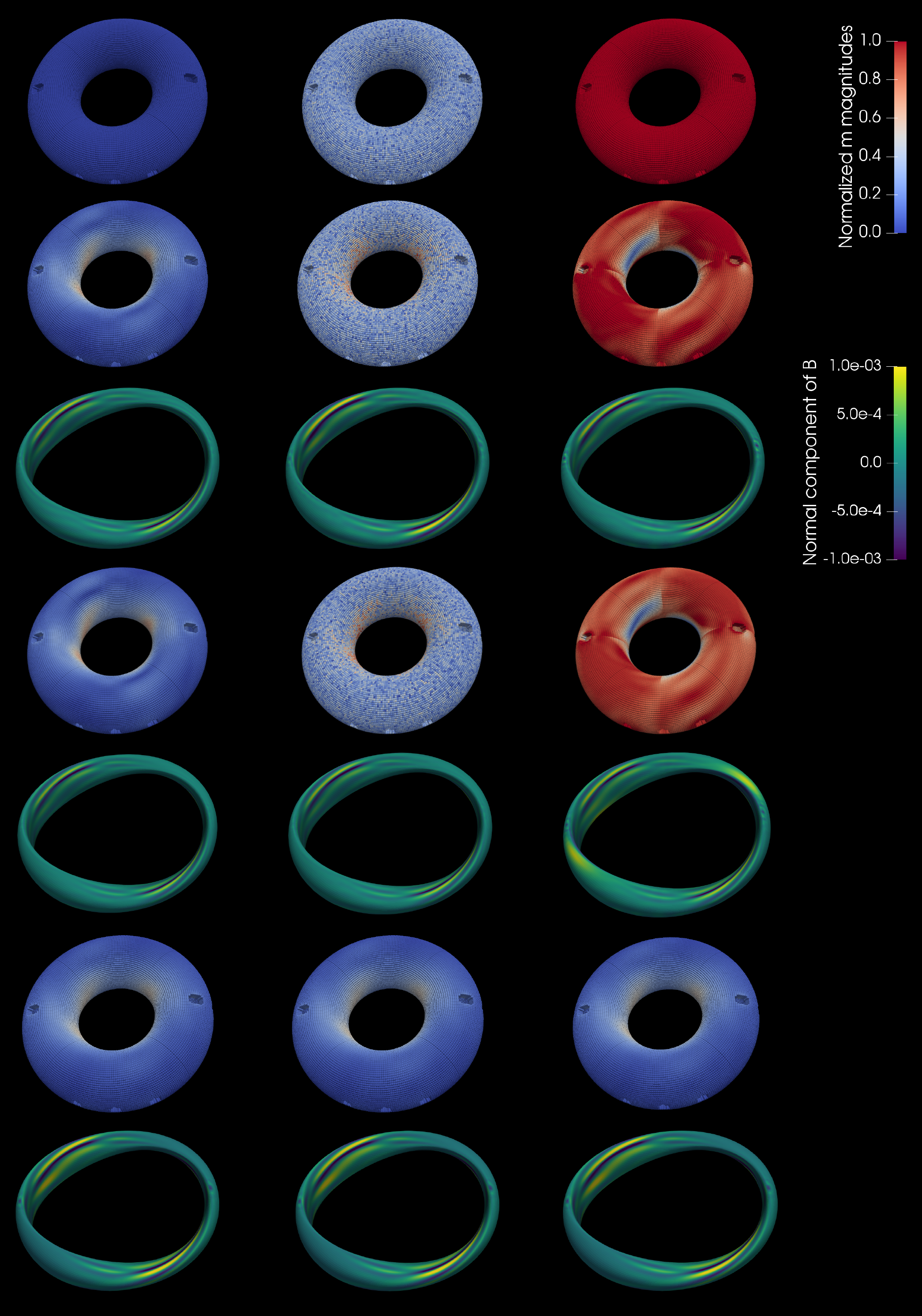}
    \put(-1, 87){\begin{rotate}{90}Initial conditions \end{rotate}}
    \put(7, 100.5){All zeros}
    \put(26.5, 100.5){All random}
    \put(48.5, 100.5){All max}
    \put(-1, 78){\begin{rotate}{90}$\lambda = 0$ \end{rotate}}
    \put(-1, 48){\begin{rotate}{90}$\lambda = 10^{-10}$ \end{rotate}}
    \put(-1, 20){\begin{rotate}{90}$\lambda = 10^{-8}$ \end{rotate}}
    \end{overpic}
    \caption{Permanent magnet manifolds (rows 1, 2, 4, 6) and $\bm B \cdot \hat{\bm n}$ on the plasma surface (rows 3, 5, 7), generated by solving \eqref{eq:convex_simple} for the MUSE permanent magnet stellarator. The toroidal field coils are omitted from the illustration. Tikhonov regularization with $\lambda = \{0, 10^{-10}, 10^{-8}\}$ and three different initial conditions were used, until a high performance value $f_B \sim 10^{-8}$ was achieved. Despite the convexity, the ill-posedness of the optimization results in vastly different permanent magnet configurations (and some variation in $\bm B \cdot\hat{\bm n}$ on the plasma surface) depending on the initial condition. Only with sufficient regularization can we break the degeneracy from the ill-posedness.}
    \label{fig:convex_illposedness}
\end{figure*}
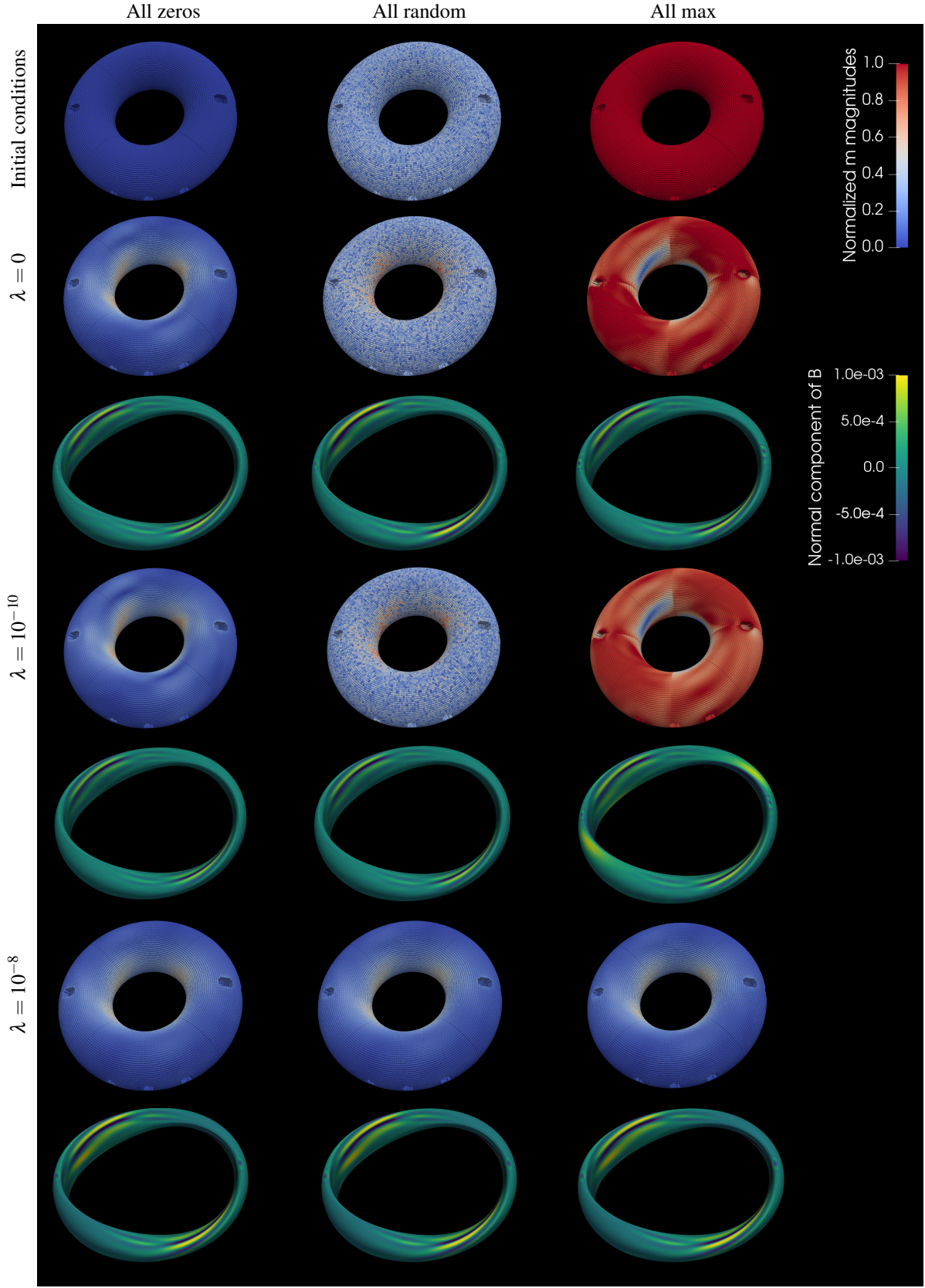

Second, even the \textit{convex} formulation of permanent magnet optimization, which omits any engineering requirements on the permanent magnets, often gives rise to different initial conditions converging to different final solutions. This pseudo-paradox is simply a result of the strong ill-posedness of the optimization problem; one should imagine the optimization landscape as a convex space but with a very large, flat valley containing many quasi-minima. It is well known that this issue is addressed with regularization. Indeed, without sufficient regularization, each new initial condition will result in a quasi-minima solution that is correct to numerical precision. This problem is so ill-posed that the quasi-minima often look like extremely different configurations of permanent magnets. Fig.~\ref{fig:convex_illposedness} shows that modest Tikhonov regularization fully circumnavigates this qualitative behavior, although the regularization may make it more difficult to access the desired parameter space of sparse, binary solutions. 

This type of ill-posedness also occurs in varieties of sparse regression. A notable example occurs in sparse system identification, where the goal is to search through very large libraries of candidate functions. In that case, the goal is to discover \textit{the} underlying equations of a dynamical system; there is a \textit{right} answer. In the context of permanent magnet optimization, this strong ill-posedness can be seen as a strength rather than a weakness since (modulo engineering constraints) there is no \textit{right} answer; \textit{any} permanent magnet configuration that produces negligible magnetic field errors on the plasma surface is a suitable solution. In other words, there are significant degrees of freedom available for incorporating additional requirements regarding the configuration of magnets. 

However, even in the permanent magnet problem there are some potential downsides of ill-posedness. Coupled with true local minima in the nonconvex setting, ill-posedness can have significant consequences regarding the number, manufacture, and cost of the permanent magnets. It also impedes conclusions about the possible parameter space of configurations; it is possible that higher-quality or easier-to-engineer configurations are available but have not been found. These problems can be compounded by a poor choice of loss terms in the objective function or because such configurations cannot be easily found with the algorithms used to minimize the objective. 

A relax-and-split algorithm is ideal for high-dimensional, constrained sparse regression.
In a relax-and-split approach, an iterative solve is set up between a convex subproblem over any set of convex constraints, and an ``easy'' nonconvex subproblem that renders the overall approach very computationally efficient.
The relax-and-split iterations continue until the convex and nonconvex subproblems both converge. The relax-and-split formulation exhibits excellent and well-understood local convergence properties~\cite{zheng2020relax}, circumnavigates the initial guess sensitivity of previous work (justified later in Sec.~\ref{sec:relax_and_split}), and can be easily extended in various ways from the baseline implementation in the open-source PySINDy code~\cite{desilva2020pysindy,Kaptanoglu2022} that is used for sparse system identification. 
Variants of relax-and-split were used in Champion et al.~\cite{champion2020unified} to solve system identification problems with affine constraints in the optimization variables, extended for additional constraints and nonconvex loss terms in Kaptanoglu et al.~\cite{kaptanoglu2021physics, kaptanoglu2021promoting}, recently used as a denoising strategy in Hokanson et al.~\cite{hokanson2022simultaneous}, and increasingly contribute to optimization problems found outside the field of system identification~\cite{zheng2021trimmed,maass2022nonconvex,sholokhov2022relaxation}. Our new implementation for permanent magnets can also be used to rapidly solve complex sparse regression problems with  $\mathcal{O}(10^4)$ $-$ $\mathcal{O}(10^6)$ unknowns and a similar number of constraints. To our knowledge, solving such large, constrained optimization problems with the $l_0$ norm has been largely infeasible with
current algorithm implementations in the sparse regression field until very recent work~\cite{bertsimas2020sparse,hazimeh2021sparse,bertsimas2022backbone}. 
By framing permanent magnet optimization as sparse regression, we build a bridge to these powerful tools.

\section{Methodology}\label{sec:methodology}
Permanent magnets can be approximated as magnetic dipoles if the distance between the field evaluation point and the center of the permanent magnet is substantially larger than the size of the permanent magnet. This is a substantial advantage for optimization because
the magnetic field generated by a magnetic dipole exhibits a simple analytic expression that is independent of the permanent magnet's geometry and linear in the dipole moment $\bm m$. Green's functions can also be used for fully calculating the near-fields of permanent magnets with simple shapes like spheres and cubes, while retaining the linearity in $\bm m$. Moreover, the dipole moment of the permanent magnet is approximately independent of the external field if the permanent magnet permeability is low. 

Together, these approximations facilitate a very simple form for the total magnetic field. It follows that a straightforward way to imagine placing permanent magnets is to approximate the magnets as magnetic dipoles, discretize a large computational domain into elements of a mesh, and represent each element as a single dipole at the center (again assuming that the element size is substantially smaller than the distance between the element and the plasma). If the total number of discrete dipoles $\bm m_i$ is $D$, then the total magnetic field $\bm B_M$ is simply a sum over all the dipoles,
\begin{align}
\label{eq:total_magnetic_dipole_field}
    \bm B_{M} = \frac{\mu_0}{4\pi}\sum_{i=1}^D\left(\frac{3\bm m_i \cdot \bm r_i}{\|\bm r_i\|_2^5}\bm r_i - \frac{\bm m_i}{\|\bm r_i\|_2^3} \right).
\end{align}
Here $\mu_0$ is the vacuum permeability, $\bm r_i$ is the vector between the evaluation point and the $i^\text{th}$ dipole, and SI units will be used throughout this work. 
Critically, Eq.~\eqref{eq:total_magnetic_dipole_field} is linear in the dipole moments $\bm m_i$. The primary target for stage-2 coil optimization is to, together with the traditional magnetic coils,  use the permanent magnets to fully match the desired magnetic fields at the plasma boundary surface, 
\begin{align}
\label{eq:f_b2_loss}
    f_B = \int \left(\left(\bm B_M + \bm B_P + \bm B_C\right)\cdot\hat{\bm n}\right)^2 d^2\bm x.
\end{align}
That is, we use the permanent magnets to minimize the value of $f_B$.
Here $\hat{\bm n}$ is the normal vector to the plasma boundary, and
$\bm B_P$ and $\bm B_C$ are the magnetic fields generated by plasma current and the traditional coils, respectively. For the purposes of this work, $\bm B_P$ is the field from a stage-1 optimized plasma boundary and $\bm B_C$ is the field generated from a minimal set of basic coils that produce the net toroidal flux. 
Eq.~\eqref{eq:f_b2_loss} is convex because it can be written as a linear least-squares term in the $\bm m_i$, as we show explicitly in Appendix~\ref{sec:appendix_least_squares}.

Next, we assume a straightforward grid of permanent magnet locations. Cartesian $(x, y, z)$, cylindrical $(r, \phi, z)$, and simple toroidal $(r_\text{minor}, \phi, \theta)$ coordinate systems are implemented in the code. Rectangular cubes, curved square bricks, and simple toroidal shells are ideal shapes for permanent magnets because they facilitate cheap, mass manufacturing and straightforward assembly. Cylindrical coordinates are used in Sec.~\ref{sec:NCSX} and simple toroidal coordinates are used in Sec.~\ref{sec:MUSE}.  Eq.~\eqref{eq:f_b2_loss} is independent of the coordinate-system choice; the coordinate system here primarily serves to specify which coordinate directions are considered grid-aligned, since this is an attractive engineering property to promote.

As in Hammond et al.~\cite{hammond2020geometric}, we define the volume as the space between two toroidal limiting surfaces. The inner surface that encloses the plasma is usually chosen to be the experimental vacuum vessel. If there is no designed vacuum vessel, a simple transformation can be used to generate an inner toroidal surface as follows. The plasma boundary normal vectors in cylindrical coordinates are projected onto the $(r$, $z)$ plane of the corresponding quadrature point (so that the new surface locations are defined at identical poloidal cross-sections), and then multiplied by an overall offset value to generate a surface for the inner toroidal boundary of the permanent magnet configuration. The outer limiting surface is generated similarly, using the projected normal vectors on the inner toroidal surface. For moderately shaped equilibria, these simple transformations work well to generate the permanent magnet volume. Any curved square bricks that are not between the inner and outer surfaces are eliminated.
Custom grids are also admissible, and future work could straightforwardly implement additional grid generation schemes.

As discussed in Sec.~\ref{sec:relationship_with_sparse_regression}, the linear least-squares part of the permanent magnet problem is strongly ill-posed. 
Typically Tikhonov regularization with strength $\lambda$ is added to the stage-2 permanent magnet optimization for regularization of $f_B$,
\begin{align}
\label{eq:Tikhonov_reg}
f_m = \lambda\sum_{i=1}^D \|\bm m_i \|_2^2,
\end{align}
and this term is also convex in the $\bm m_i$. A downside of this term is that it tends to select for solutions with weak but nonzero magnets. For more general regularization terms, we consider,
\begin{align}
    f_m = \lambda \sum_{i=1}^D R(\bm m_i),
\end{align}
where the sum over $i$ could also be chosen to be inside the regularizer $R$. Here $R(\bm m_i)$ is smooth and convex, such as the commonly used $l_2$ norm; nonsmooth and nonconvex regularizers are discussed below but we denote these separately.

\subsection{Optimization objectives}
Before formulating the full optimization problem to solve, it is illustrative to be explicit about the goals of the optimization.
Consider the following objectives that we would like to promote or impose in the optimization problem,
\begin{enumerate}
    \item Fit the MHD equilibrium fields: minimize $f_B$. \label{item1}
    \item Regularize $f_B$: minimize $f_m$. \label{item2}
    \item Avoid magnetic dipole values above the maximum magnetization of the material: impose the hard constraints that $\|\bm m_i\|_2 \leq m_i^\text{max}$ where the maximum value for dipole moment $i$ is defined through the remanence field $B_\text{rem}$ and the cell volume $V^\text{cell}_i$, $m_i^\text{max} = \mu_0^{-1}B_\text{rem}V^\text{cell}_i$. For this work, we assume $B_\text{rem} = 1.465$ T to match the value used for FAMUS runs of the MUSE stellarator. MUSE uses commercial ``N52'' neodymium-iron-boron (Nd2Fe14B)  magnets.
    \label{item3}
     \item Produce a sparse, binary solution. Bias the dipoles towards either satisfying $\|\bm m_i\|_2 = 0$ or $\|\bm m_i\|_2 = m_i^\text{max}$ for all $i$, so that permanent magnets are either omitted or placed with approximately maximum strength. Moreover, minimize the cost of the permanent magnetic configuration by using as few magnets as possible. \label{item4}
    \item Promote properties that reduce the engineering complexity of the magnet configuration, such as constraining each dipole's direction to be ``grid-aligned'', i.e. perpendicular to the face of each grid element. \label{item5}
\end{enumerate}
Below, we start with a convex formulation that is sufficient for objectives~\ref{item1}$-$\ref{item3} in Sections~\ref{sec:first_algorithm}$-$\ref{sec:MwPGP}, and then add in  nonconvexity for promoting sparse configurations of permanent magnets that address objectives~\ref{item4}$-$\ref{item5} in Sections~\ref{sec:nonconvex_loss_terms}$-$\ref{sec:relax_and_split}.

\subsection{A first attempt at formulating the optimization}\label{sec:first_algorithm}

We now propose a first objective function that accomplishes the optimization objectives~\ref{item1}$-$\ref{item3}. First, define the vector of optimization variables, $\bm m = [m^x_1, m^y_1, m^z_1, m^x_2, ..., m^z_D]$. Then the proposed objective function can be formulated,
\begin{align}
\label{eq:convex_simple}
    &\argmin_{\bm m}\frac{1}{2}\|\bm A \bm m - \bm b\|^2_2 + \lambda R(\bm m) \\ \notag
    &\|\bm m_i \|_2^2 \leq \left(m_i^\text{max}\right)^2, \quad i=1,...,D.
    % \quad (m^x_1)^2 + (&m^y_1)^2 + (m^z_1)^2 \leq \left(m^\text{max}_1\right)^2 \\ \notag
    % &\vdots \\ \notag
    % \quad (m^x_D)^2 + (&m^y_D)^2 + (m^z_D)^2 \leq \left(m^\text{max}_D\right)^2.
\end{align}
The quadratic constraints are convex since each $\bm m_i$ vector must be contained in a $l_2$ ball in $\mathbb{R}^3$. For the remainder of the work, we will denote as $\mathbb{S}_m$ the hypersurface spanned by the intersection of the spherical constraints. Critically for efficiency, the constraints are separable. Unlike most stellarator optimization problems, we do not convert these constraints into a related loss term -- we impose them as hard constraints. Interior point methods (IPMs) may also be used here, where the inequality constraints would be converted into barrier functions in the loss terms, but IPMs usually do  not scale very well to high-dimensional settings. Fortunately, for convex problems with a large number of separable, convex constraints, there are other fast algorithms.

Before moving on to an algorithm for Eq.~\ref{eq:convex_simple}, we note here that the demagnetization effects from finite permanent magnet coercivity could also be modeled as convex constraints. These constraints could be critical for accurate modeling of stellarators with large magnetic field strengths. If the dipole at position $\bm x_i$ demagnetizes if the external field strength is greater than $B_\text{max}$, this can be posed as 
\begin{align}
    \|\bm B_M(\bm m, \bm x_i) + \bm B_P(\bm x_i) + \bm B_C(\bm x_i)\|_2^2 \leq B_\text{max}^2,  \label{eq:coercivity_constraint}
\end{align}
It should be clear that the field from the $i$th dipole is excluded from  expression~\eqref{eq:coercivity_constraint} when evaluated at $\bm x_i$, and that for permanent magnets that are very close to the $i$th permanent magnet, there are significant errors introduced by the dipole approximation (however, these errors can be avoided with spherical permanent magnets or accounted for if other simple shapes are used). 
There are constraints of the  Eq.~\eqref{eq:coercivity_constraint} type from all the dipole locations, adding another $D$ constraints to the values in $\bm m$ if each constraint is active.
Each constraint is a convex but not separable constraint on the $\bm m_i$ and could be implemented with barrier functions in future work.

\subsection{An algorithm for simple permanent magnet optimization}\label{sec:MwPGP}
The optimization problem in~\eqref{eq:convex_simple} is a convex problem with separable, convex, spherical constraints. Therefore, it can be solved with many different algorithms, and the global minimum can be found. The primary issues are the high-dimension of the problem and the large number of constraints.
The problem in~\eqref{eq:convex_simple} forms the backbone of the present work, so we first provide an algorithm that can efficiently solve this problem despite the high dimension.

Many algorithms have been considered for quadratic programs with quadratic constraints~\cite{an2000efficient}. It is only recently that fast solvers have been devised for exactly the case of high-dimensional convex optimization problems with a large number of spherical constraints~\cite{dostal2016optimal}. In particular, an algorithm is proposed in Bouchala et al.~\cite{bouchala2014solution} that can optimally solve these problems, and we adopt this algorithm for our purposes. An extended algorithm and other variations~\cite{pospivsil2018projected} are also available if additional equality constraints are present. The algorithm is essentially projected-gradient descent (the standard algorithm for high-dimensional, smooth, constrained optimization) with a conjugate-gradient acceleration, and as such, it relies on a explicit form for the projection operator,
\begin{align}
    P_{\mathbb{S}_m}(\bm m) = \argmin_{\bm y \in \mathbb{S}_m}\|\bm m - \bm y\|^2_2.
\end{align}
The projection can be decomposed as,
\begin{align}
    P_{\mathbb{S}_m}(\bm m) = [P(\bm m_1), ..., P(\bm m_D)],
\end{align}
by the separability of the constraints, and each of the projections satisfies,
\begin{align}
    P(\bm m_i) = \argmin_{\|\bm y\|^2_2 \leq \left(m_i^\text{max}\right)^2}\|\bm m_i - \bm y\|^2_2
    = \frac{\bm m_i}{\max\left(1, \frac{\|\bm m_i\|_2}{m^\text{max}_i}\right)}.
\end{align}
Lastly, we typically initialize the algorithm with an initial guess for the dipoles as all zeros. The natural guess is $\bm m^{(0)} = P_{\mathbb{S}_m}(\bm A^\dagger \bm b)$, where $\bm A^\dagger$ denotes the pseudo-inverse of $\bm A$, but we found that this often generates very poor initial guesses when the constraints are very active. Alternatively, when the problem is strongly ill-posed, it is often useful to start with the dipoles at maximum strength to preference the algorithm towards solutions with strong magnets. Regardless, the algorithm requires that $\bm m^{(0)} \in \mathbb{S}_m$. Many other initial conditions were tested to verify that, with sufficient regularization, the algorithm correctly converges to the minimum of this convex problem, as in Fig.~\ref{fig:convex_illposedness}.

\subsection{Sparse optimization for permanent magnet design}\label{sec:nonconvex_loss_terms}

Unfortunately, the formulation in~\eqref{eq:convex_simple} is not sufficient for a sparse, binary solution. These are attractive properties to promote, since it means only a sparse collection of maximum-strength permanent magnets can be used. We begin this section with a discussion of one of the foundations of sparsity-promoting optimization: the $l_0$ ``norm'', which is simply an operator that counts up the number of nonzero terms in a vector. The $l_0$ norm has not been used before for permanent magnet optimization because it appears very challenging to address with traditional optimization. For instance, the $l_0$
 norm is nonsmooth, preventing the straightforward application of generic algorithms for nonconvex optimization based on gradient and Hessian information such as BFGS~\cite{liu1989limited}. The high-dimension of the problem and large number of constraints further reduces the number of fast algorithms that are suitable for addressing objective functions with the $l_0$ norm. Nonetheless, we define a new objective function with the $l_0$ norm and borrow recent ideas from the field of sparse regression to effectively solve it,
\begin{align}	
\label{eq:L0}
    \argmin_{\bm m\in\mathbb{S}_m}&\frac{1}{2}\|\bm A \bm m - \bm b\|^2_2 + \lambda\| \bm m\|_2^2  + \alpha\|\bm m \|_0.
\end{align}
We will show below that the formulation in~\eqref{eq:L0} now satisfies all of the requirements for the permanent magnet optimization problem and the partial convexity of the objective function can be utilized.
Moreover, the objective function in~\eqref{eq:L0} can be altered without significant changes in procedure; substitutions for the $l_0$ norm in~\eqref{eq:L0} can be made for other regularizers, depending on the optimization goals. Examples include the $l_1$, group $l_{2, 1}$, and group $l_{0, 1}$ norms. For instance, the \textit{non-overlapping} group $l_{0,2}$ or $l_{0,1}$ operators~\cite{matsuoka2017joint} could be used to enforce an $l_0$ norm-like structure on the magnitudes of each dipole, allowing for dipoles that do not align with any of the grid directions$^*$\blfootnote{$^*$ Consider a set of subsets $\{G_{1}, G_{2}, ...\}$ such that the union of the subsets is the entire index set of the optimization variables, $G_{1} \cup G_{2} \cup \cdots = \{1, ..., D\}$. The group $l_{0, 1}$ norm is defined as $\|\|\bm m\|^{G_1}_1, \|\bm m\|_1^{G_2}, ..., \|_0$. For overlapping groups, this can be a very complicated operator. For the permanent magnet problem, the groups are the non-overlapping triplets consisting of the components of each dipole vector. In the non-overlapping case, these operators can be addressed like the $l_0$ is addressed in Appendix~\ref{sec:appendix_prox}.}. These operators can also be used in the algorithm proposed below but we focus on the $l_0$ norm in this work both for concreteness and because it satisfies all the requirements for generating sparse, high-quality configurations of permanent magnets. As we detail our algorithm, it should become apparent that the $l_1$ loss term, the typical choice for a relaxation of the $l_0$ loss term, is not sufficient for generating sparse and \textit{binary} solutions. 

Beyond the connection with sparse regression, it may not be clear yet why this formulation of the permanent magnet optimization problem could be preferable to formulations proposed in previous works. In Section~\ref{sec:relax_and_split} we will show that we can use the special structure appearing in our formulation to design a  fast, flexible, easy-to-understand solution to placing a sparse set of permanent magnets for stellarator field shaping.

\subsection{Proposed relax-and-split algorithm}\label{sec:relax_and_split}
An algorithm is required for effectively solving the various proposed objective functions that can be used for permanent magnetic optimization. The backbone of all these objective functions is a linear least-squares problem with a convex regularizer, subject to a large number of separable, spherical constraints. 
Nonconvex optimization problems with convex constraints can be effectively solved with the relax-and-split formulation~\cite{zheng2020relax} of the optimization problem if the proximal operator,
\begin{align}
    \mathrm{prox}_{\alpha N}(\bm m) \equiv \argmin_{\bm y}\left[\frac{1}{2}\|\bm m - \bm y \|_2^2 + \alpha N(\bm y)\right],
\end{align}
of the nonconvex (and/or nonsmooth) part of the objective function, $N(\bm m)$, is known or easily approximated. In the present work, it is assumed that an analytic expression is known for the proximal operator (as is the case for the $l_0$, $l_1$, non-overlapping group $l_{0, 1}$, etc.), which significantly simplifies our task and additionally implies that, despite the notation, $N(\bm m)$ is not an arbitrary function. There are also nonconvex loss terms without known analytic proximal operators that nonetheless can be rapidly numerically computed. 

To see how the proximal operator enters into the optimization, we first introduce a proxy variable $\bm w \sim \bm m$. This proxy is kept reasonably close to the values in $\bm m$ by the inclusion of a new least-squares term $\nu^{-1} \|\bm m - \bm w\|_2^2$, with the value of the hyperparameter $\nu$ determining how closely the two variables should match. The goal is now to solve the following optimization problem,
\medmuskip=-1mu
\thinmuskip=0mu
\thickmuskip=0mu
\begin{align}
    \label{eq:general_relax_and_split}
    \argmin_{\bm w,\bm m\in\mathbb{S}_m}&\left[\frac{\|\bm A \bm m - \bm b\|^2_2}{2} + \frac{\| \bm m - \bm w\|^2_2}{2\nu} + \lambda R(\bm m)\right] + \alpha N(\bm w).
\end{align}
We can solve this problem effectively by using variable projection~\cite{van2021variable}. We solve the inner optimization problem over $\bm m$ at fixed $\bm w$, then solve the outer optimization over $\bm w$ at fixed $\bm m$, and iterate between these solves until convergence is found.
The inner optimization problem is convex over the original dipole variables in $\bm m$. Fix the initial guess for $\bm w^{(0)}$ and denote the solution after the first convex iteration 
\begin{align}
    \bm {m}^{(1)} \equiv \argmin_{\bm m\in\mathbb{S}_m}\left[\frac{\|\bm A \bm m - \bm b\|^2_2}{2} + \frac{\| \bm m - \bm w^{(0)}\|^2_2}{2\nu} + \lambda R(\bm m)\right],
\end{align}
\medmuskip=4mu
\thinmuskip=4mu
\thickmuskip=4mu
so that the remaining optimization problem is
\begin{align}
    \label{eq:relax_and_split_first_iteration}
    \bm w^{(1)} &\equiv \argmin_{\bm w }\left[\frac{1}{2\nu}\| \bm m^{(1)} - \bm w\|^2_2 + \alpha N(\bm w)\right] \\ \notag &=  \mathrm{prox}_{\nu\alpha N}(\bm m^{(1)}).
\end{align}
In other words, the remaining optimization problem is equivalent to computing the proximal operator. The proximal operator for the $l_0$ norm is \textit{hard-thresholding}, detailed briefly in Appendix~\ref{sec:appendix_prox}. The hard thresholding naturally generates sparse vectors. Now we can iterate between the convex and nonconvex optimization subproblems so that the full algorithm can be summarized in Algorithm~\ref{algo}. 

\begin{algorithm}[H]
	\caption{Relax-and-split with the $l_0$ loss term}\label{algo}
	\begin{algorithmic}
		\INPUT{Magnet geometry encoded in $\bm A$, coil and plasma magnetic fields in $\bm b$, initial guess $\bm w^{(0)}$, error tolerance $\epsilon$, initial hard threshold $\alpha_0$, and strength of any additional regularization via $\lambda$.}
		\OUTPUT{Solutions $\bm m^*$ and $\bm{w}^{*}$}.  
		\Procedure{Relax-and-split}{$\bm A$, $\bm b$, $\lambda$, $\nu$, $\alpha_0$}
		\State for $\alpha = \alpha_0,2\alpha_0,...$
		\State $\quad$ for $k = 0,...,k_{max}$
		\medmuskip=0mu
		\thinmuskip=0mu
		\thickmuskip=0mu
		\begin{align}\notag\quad\quad\quad\quad\text{    } \bm{m}^{(k+1)} = \argmin_{\bm{m}\in\mathbb{S}_m}\left[\frac{\|\bm A \bm m - \bm b\|^2_2}{2} + \frac{\| \bm m - \bm w^{(k)}\|^2_2}{2\nu} + \lambda R(\bm m)\right]\end{align},
		\medmuskip=4mu
		\thinmuskip=4mu
		\thickmuskip=4mu
		\State $\quad\quad \bm{w}^{(k+1)} = \mathrm{prox}_{\nu \alpha\|(.)\|_0}(\bm m^{(k+1)})$
		\State $\quad\quad \text{if}\,\, \|\bm w^{(k+1)} - \bm w^{(k)}\|_2 \leq \epsilon$ 
		\State $\quad\quad\quad$ break
		\State \quad $\bm w^{(0)} = \bm w^{(k + 1)}$
    \State return $\bm m^{(k+1)}, \bm w^{(k+1)}$
		\EndProcedure
	\end{algorithmic}
	In other words, for each value of the hard threshold $\propto \sqrt{\alpha_0}$, solve the convex part of the optimization, take the proximal operator of the resulting vector, and iterate until convergence is found. Typically $\nu \sim \|\bm A\|_2^{-2}$ works well, so $\lambda$ and $\alpha$ can be considered the primary hyperparameters. $\lambda$ tunes the strength of the regularization and $\alpha$ tunes the strength of the nonconvexity.
\end{algorithm}

We found empirically that it often works very well to start with weak values of the hard threshold used in the $l_0$ norm (the threshold is proportional to $\sqrt{\alpha}$), converge this problem, and then use the solution as an initial guess to a new problem with a larger value of $\alpha$. This is repeated until $\alpha$ is large enough to threshold off essentially all of the magnets that are not maximum strength. Since the magnets are all essentially maximum strength by the final optimization loop, they are also all grid-aligned by virtue of the $l_0$ loss term. To see this, note that the algorithm thresholds off components of each $\bm m_i$ that are below some minimum strength, subject to the constraints on the maximum dipole magnitudes. For instance, if components of $\bm m_i$ with below $97.5\%$ of the maximum strength are thresholded, it follows that this dipole must either be exactly zero or have a single nonzero vector component. As far as we are aware, this iterative approach has not been used before in the field of sparse regression.

There are some significant advantages of Algorithm~\ref{algo}. The convex and nonconvex parts of the optimization problem can both be efficiently solved, additional convex equality and inequality constraints can be added straightforwardly, and the requirements on $\bm m$ are split between $\bm m$ and the proxy variable $\bm w$, allowing the algorithm to ``relax'' into satisfying the constraints. The formulation as sparse regression allows for the rapid adoption of concepts, algorithms, and expertise from the prolific field of sparse regression. 

Moreover, if the problem is reasonably well-posed (e.g. Tikhonov regularized), the initial permanent magnet configuration is unimportant $-$ whatever it is, it is erased by the convex optimization in the first step, fully circumnavigating part of the initial guess issues present in the other nonconvex algorithms for permanent magnet optimization. It may appear at first that this insensitivity to the initial condition may cause the algorithm to only explore a reduced space of configurations. But we now also have additional hyperparameters, and by sweeping the value of these hyperparameters, we have a systematic way to explore a very wide space of permanent magnet configurations.

\section{Results}\label{sec:Results}
We now present optimized permanent magnet configurations for the NCSX and MUSE stellarators. Both examples are run with a set of high-resolution quadrature points on the unique part of the plasma boundary $-$ 64 points in $\theta$, and 64 points
in $\phi$ ($\times 2N_{fp}$ for all $N_{fp}$ field periods). Both of the following considered stellarators are stellarator-symmetric and field-period symmetric, so it is only required to design dipoles for $1/ 2 N_{fp}$ of the toroidal angle extent, and then to repeat this configuration around the torus. 

\subsection{NCSX}\label{sec:NCSX}

\begin{figure*}
    \centering
    \begin{overpic}[width=0.94\linewidth]{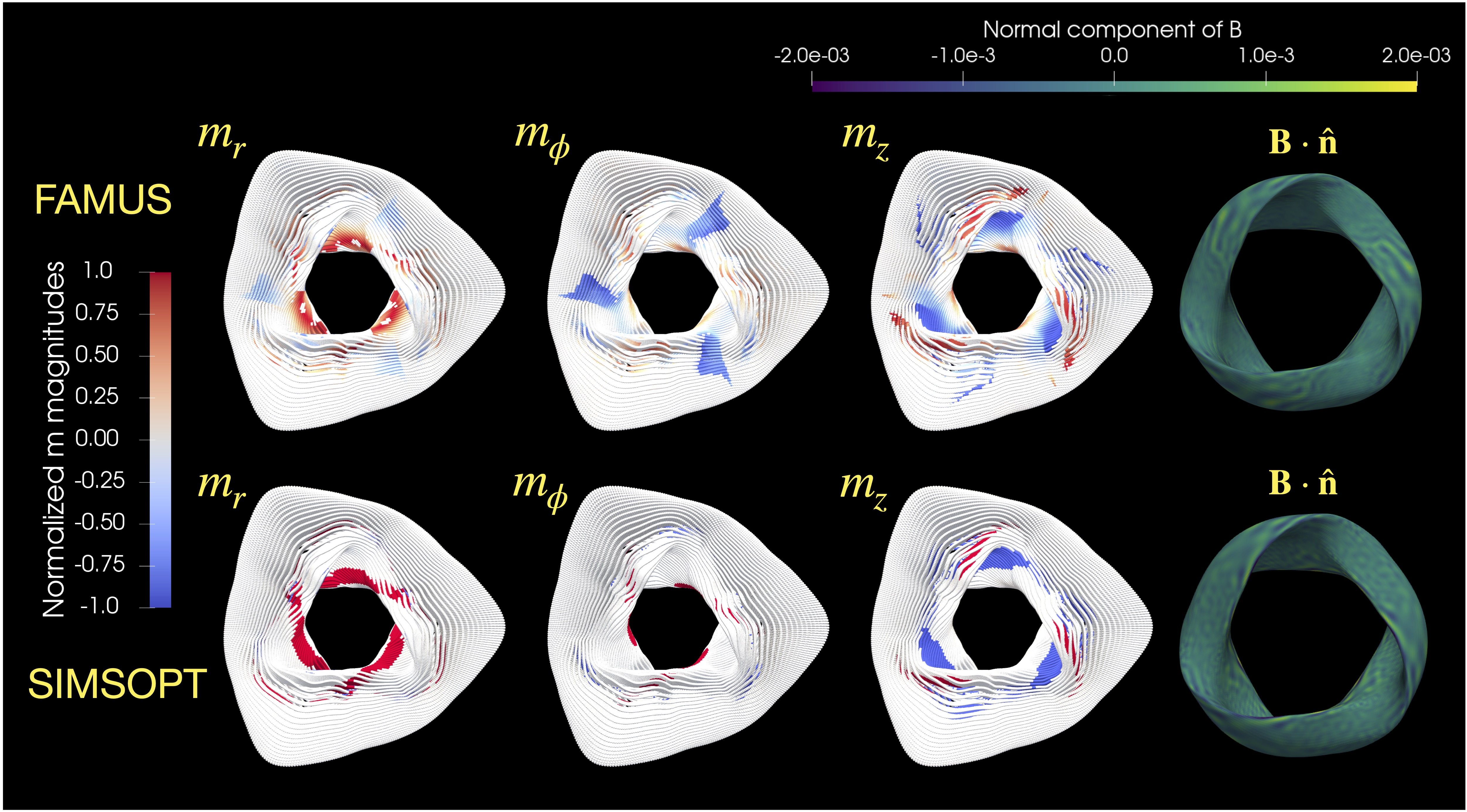}
    \end{overpic}
    \caption{Comparison between FAMUS and relax-and-split solutions on an example NCSX permanent magnet manifold. Note that the relax-and-split dipoles are aligned with the cylindrical grid.}
    \label{fig:NCSX_comparison}
\end{figure*}

NCSX was a planned quasi-axisymmetric stellarator that was partially built at the Princeton Plasma Physics Laboratory. It was originally designed with 18 modular coils and 18 planar coils. The equilibrium of interest, C09R00, was scaled to have an on-axis
magnetic field strength of $0.5$ T, which is the maximum field produced by the existing planar coils. C09R00 also exhibits a three-fold field symmetry, major radius of $1.44$ m, minor radius of $0.32$ m, and volume-averaged plasma beta $\langle \beta\rangle = 4.09\%$. However, for direct comparison with the FAMUS solution in~\cite{landreman2021calculation}, the C09R00 shape is used but with no plasma current, i.e. $\langle \beta\rangle = 0$, and the toroidal field was taken to be perfectly toroidal with no ripple.

For a direct comparison with the FAMUS~\cite{zhu2020topology} method, 
Algorithm~\ref{algo} is used in a cylindrical coordinate system with the $l_0$ norm and Tikhonov regularization is omitted. Tikhonov regularization was simply not required for a high-quality solution (and the relax-and-split term, $\propto \|\bm m - \bm w\|_2^2$, is an additional source of regularization).

The FAMUS grid of allowed dipole locations has a
resolution of 14 points radially,  so $\bm A \in \mathbb{R}^{4096 \times 172032}$. We use the same grid for our optimizations in SIMSOPT. FAMUS is used with a level of regularization for a plausible solution 
$f_B \approx 1.6 \times 10^{-6}$ T$^2$m$^2$. An effective volume of the permanent magnet region can be defined by 
\begin{align}\label{eq:eff_vol}
V_\text{eff} = \sum_{j=1}^D \frac{\|\bm m_j\|_2}{M_0}, \quad M_0 \equiv \frac{B_\text{rem}}{\mu_0},
\end{align}
and the result is 2.29 m$^3$ out of a maximum available volume of $3.23$ m$^3$. 
To measure how binary a solution is, we also define the binary fraction
\begin{align}\label{eq:B_delta}
    f_\delta = 1 - \frac{\#\{i | \delta \leq \|\bm m_i\|_2 \leq 1-\delta\}}{D}.
\end{align}
There is no post-processing optimization done to the FAMUS solution, so there is significant ``pile-up'' near dipole magnitudes at 0 and 1 and $f_{0.01} = 0.57$.

\begin{figure}
    \centering
    \begin{overpic}[width=0.97\linewidth]{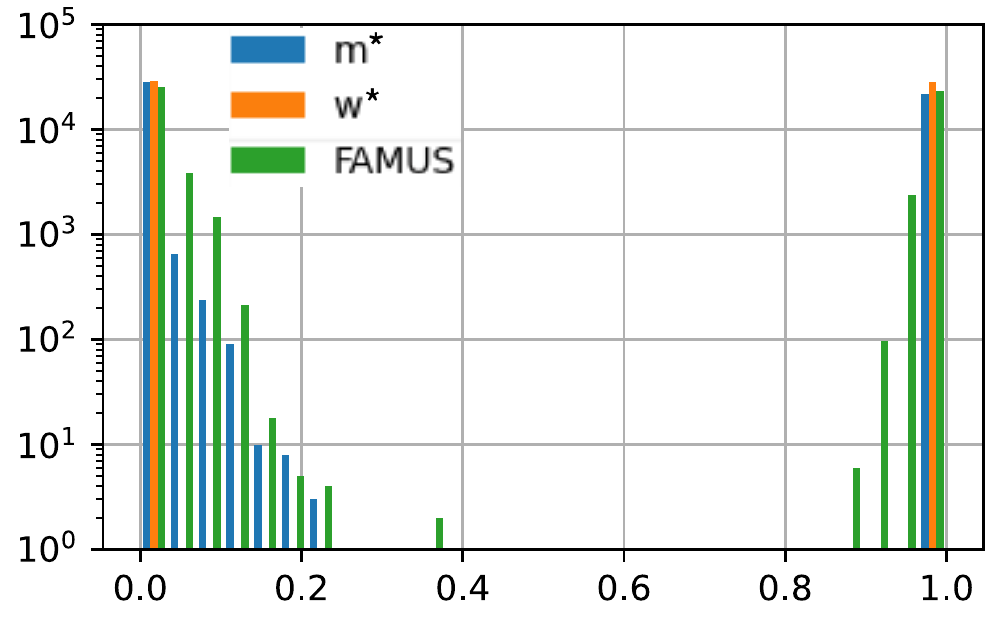}
    \put(-0.5, 15){\begin{rotate}{90}Number of magnitudes \end{rotate}}
    \put(24, -3){Value of $\|\bm m_i\|_2 / m_i^\text{max}$ for all $D$ dipoles}    \end{overpic}
    \vspace{0.1in}
    \caption{Distributions of the dipole magnitudes for the relax-and-split solutions and FAMUS. Significant numbers of weak magnets seem to be required for low-error configurations with this NCSX permanent magnet manifold.}
    \label{fig:ncsx_histograms}
\end{figure}

A representative SIMSOPT $\bm m^*$ solution achieves a similar $f_B \approx 1.6 \times 10^{-6}$ T$^2$m$^2$ with a very similar effective volume of $2.34$ m$^3$ and a substantially improved $f_{0.01} = 0.84$. This improvement with SIMSOPT occurs despite the fact that the relax-and-split method also produces a cylindrically grid-aligned solution. This solution is compared with the FAMUS solution in Figures~\ref{fig:NCSX_comparison} and \ref{fig:ncsx_histograms}. The $\bm w^*$ solution achieves a poor value of $f_B$, but nonetheless still successfully pulls $\bm m^*$ towards a high-quality solution that is approximately sparse. The relax-and-split and FAMUS solutions have some qualitative similarities in the strength of the permanent magnets. It is noteworthy that in Fig.~\ref{fig:ncsx_histograms}, both solutions appear to require a set of weak magnets for low-error configurations of this NCSX example. This lends some evidence to the following propositions: that the permanent magnet grid choice can be reasonably important for finding high-performance solutions, and that the low-errors in $f_B$ are quite sensitive to small changes in the magnets. Overall, the major takeaway is that we have found a relax-and-split solution that provides similar $f_B$ performance as the FAMUS solution, despite being significantly sparser and grid-aligned.

\begin{figure*}[t]
    \centering
    \begin{overpic}[width=0.9\linewidth]{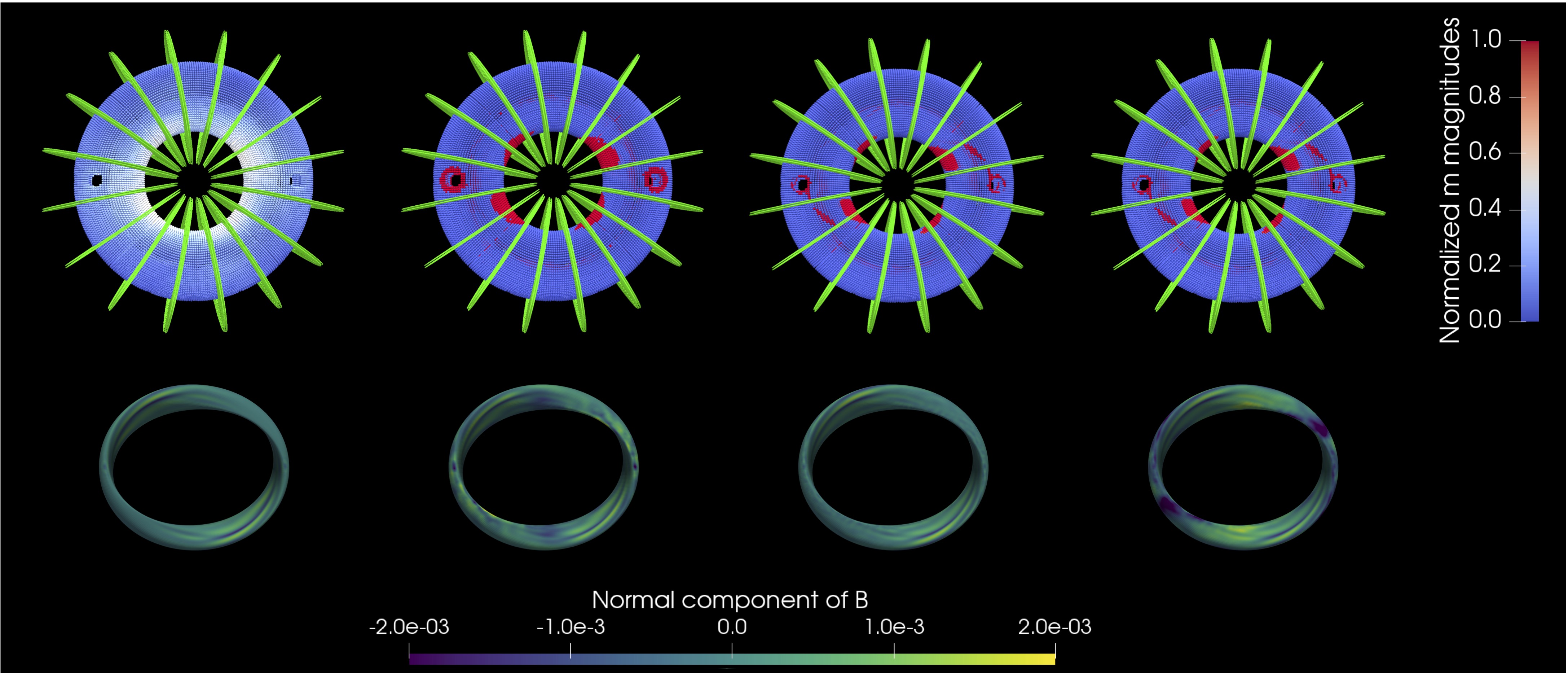}
    \put(9, 44){Convex}
    \put(32, 44){FAMUS}
    \put(56, 44){$\bm m^*$}
    \put(79, 44){$\bm w^*$}
    \end{overpic}
    \caption{Comparison between the convex algorithm, the FAMUS solution, and the relax-and-split solutions for MUSE. The sparse relax-and-split $\bm w^*$ produces slightly larger $f_B$ error than FAMUS, but uses fewer magnets.}
    \label{fig:muse_summary}
\end{figure*}

\begin{figure*}[ht]
    \centering
    \includegraphics[width=0.74\linewidth]{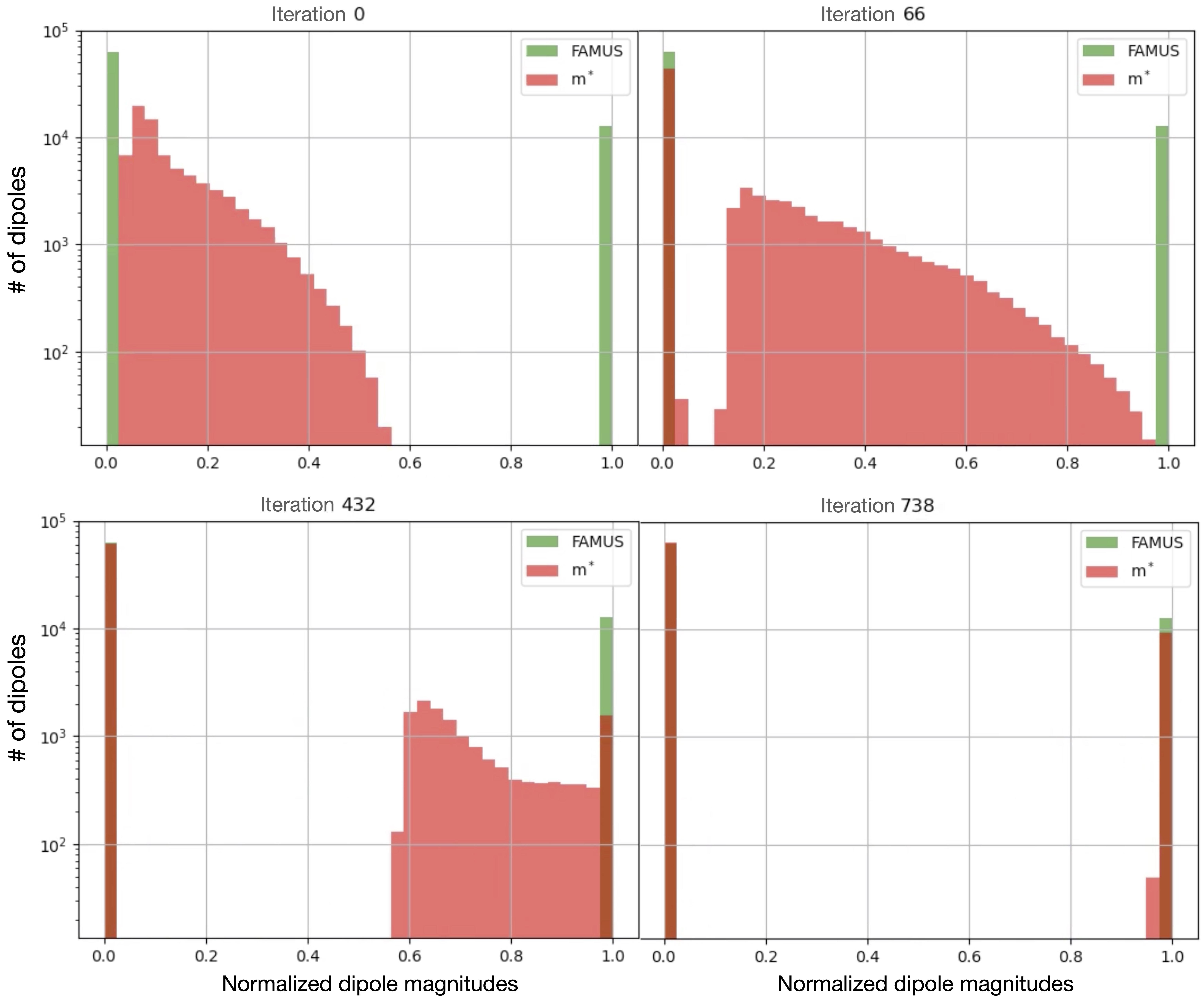}
    \caption{Histograms of the dipole magnitudes as the algorithm progresses (red) compared with the optimized FAMUS solution for MUSE (green). Slowly increasing the thresholding has the effect of pushing all the magnets to magnitudes of zero or one. }
    \label{fig:histogram_progress}
\end{figure*}

\subsection{The MUSE stellarator}\label{sec:MUSE}
MUSE is a table-top stellarator experiment using permanent magnets that is currently under construction~\cite{qian2022simpler}. MUSE was optimized for a high degree of quasi-symmetry, and the experiment's permanent magnet configuration was optimized to have good flux surfaces with the FAMUS code~\cite{qian2021stellarator} to $f_B \approx 5.17\times 10^{-8}$ T$^2$ m$^2$. The dipoles were constrained to point only in the minor radial direction in simple toroidal coordinates. Diagnostic ports and other spatial restrictions are used to represent the real permanent magnet configuration. Spatial restrictions only eliminate a portion of the grid from consideration, and importantly do not change the properties of the optimization problem.
After the permanent magnets are optimized, the combined magnetic field from the magnets and coils are used to compute Poincar\'{e} plots in order to verify the quality of the flux surfaces. Substantial discrete optimization was used in post-processing steps to achieve engineering constraints while preserving the physics objectives, as described in Qian et al.~\cite{qian2022simpler}. 

We use the same stage-1 optimized plasma surface, the same 16 planar TF coils, and the same MUSE permanent magnet grid array of four toroidal quadrants~\cite{qian2022simpler}  (i.e. the dipole locations are identical in the FAMUS and SIMSOPT optimizations) with a simple toroidal coordinate system. The grid provides for the inclusion of four vertical ports and twelve horizontal ports. We solve both the convex and nonconvex optimization problem from~\eqref{eq:convex_simple} and~\eqref{eq:L0} and compare with the FAMUS solution. For these optimizations, the number of dipoles is 75,460 ($\times 4$ via symmetries), each with three vector components, so $\bm A \in R^{4096\times 226380}$. Tikhonov regularization is used with strength $\lambda = 10^{-8}$ for the convex problem; these results are identical to the last two rows of Fig.~\ref{fig:convex_illposedness}. Tikhonov regularization is omitted for the SIMSOPT solution with the $l_0$ term included, for the same reason as in Sec.~\ref{sec:NCSX}. 

The results comparing the  convex optimization in~\eqref{eq:convex_simple}, the FAMUS solution, and the relax-and-split solutions $\bm m^*$ and $\bm w^*$ are shown in Fig.~\ref{fig:muse_summary}. The toroidal field coils are shown in green, the normalized permanent magnet strength is indicated in the blue-red colorbar, and the normal component of the residual field errors is illustrated in the purple-yellow colorbar.
\begin{figure}
    \centering
    \includegraphics[width=0.95\linewidth]{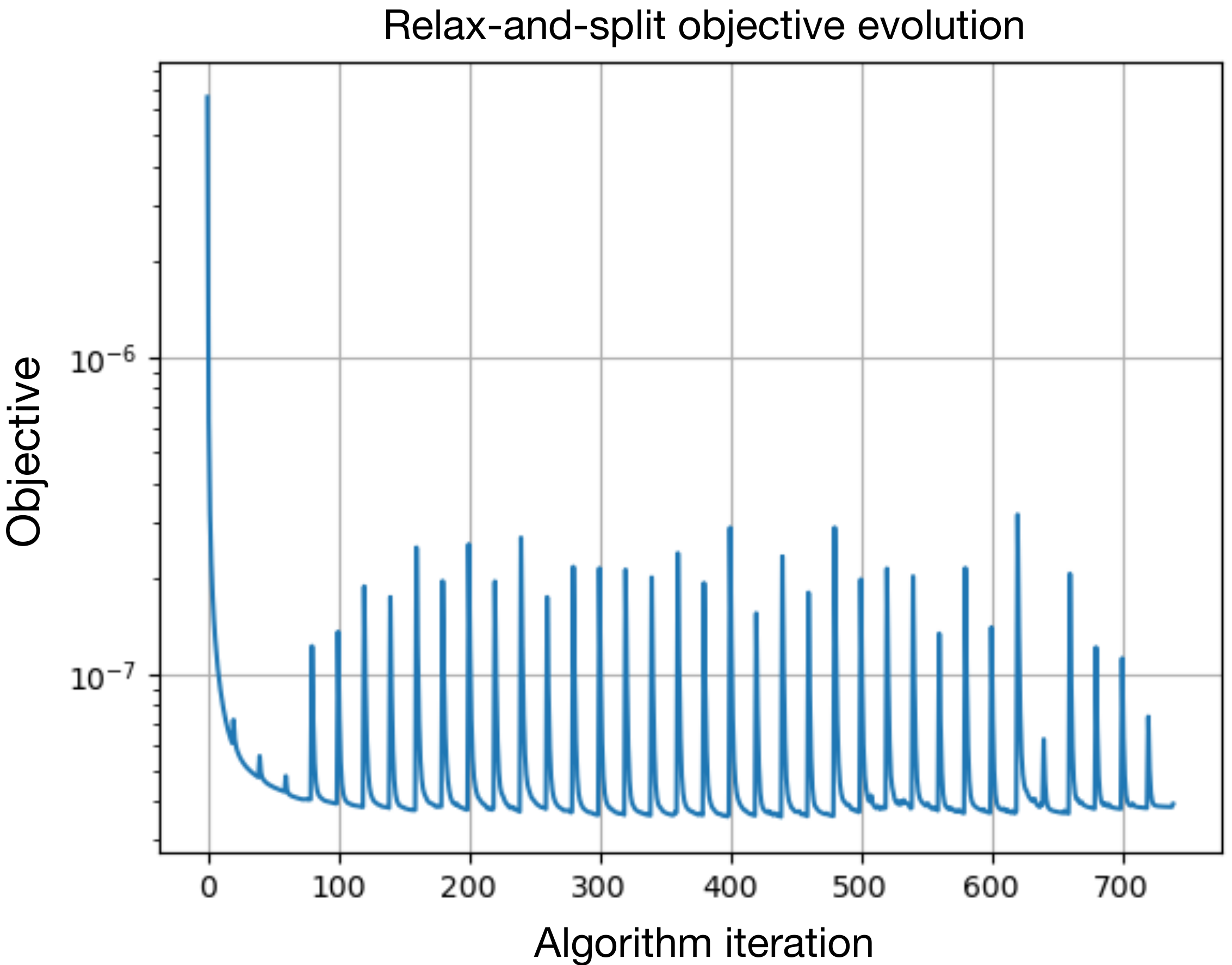}
    \caption{Illustration of the optimization objective as the relax-and-split algorithm progresses.  Increases in the hard threshold causes sudden spikes in the objective value, but the algorithm quickly recovers to pre-thresholding error levels. At the end of the algorithm, when only magnets of strength $97.5\%$ remain in the MUSE permanent magnet manifold, the error is roughly equal to the error at the beginning, when dipole magnitudes could  vary between $0$ and $1$.}
    %\vspace{-0.3in}
    \label{fig:algorithm_progress}
\end{figure}
\begin{figure}
    \centering
    \begin{overpic}[width=0.95\linewidth]{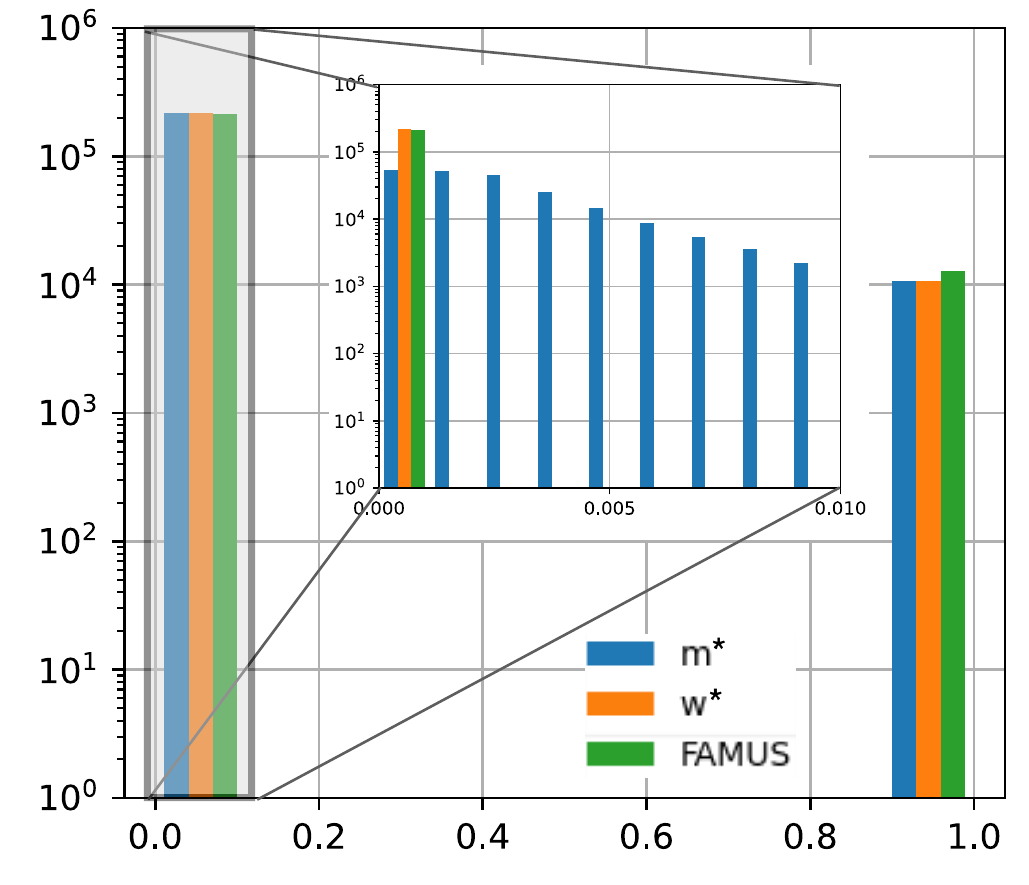}
    \put(1, 15){\begin{rotate}{90}Number of optimization variables \end{rotate}}
    \put(13, -1){Value of $|m_i|$ for all $3 \times D$ optimization variables}
    \end{overpic}
    \caption{Distributions of all the permanent magnet optimization variables for the relax-and-split and FAMUS solutions for MUSE. The vast majority of the dipoles are zeroed and relax-and-split uses fewer full-strength magnets. A zoomed view illustrates the magnet ``stragglers'' in $\bm m^*$, while $\bm w^*$ \textit{fully} zeroes out magnets below the hard threshold.}
    \vspace{-0.12in}
    \label{fig:muse_histograms}
\end{figure}
The optimization began by truncating magnets with strengths below $5\%$ and the threshold parameter (proportional to $\alpha_0$) was increased until magnets below $97.5 \%$ strength were truncated. The algorithm progress is visualized in Figures~\ref{fig:histogram_progress} and \ref{fig:algorithm_progress}, showing that the optimization successfully transforms a nonsparse solution of weak dipoles to a very sparse solution of maximum-strength dipoles.

The $\bm m^*$ solution achieves $f_B \approx 2.8
\times 10^{-8}$ and the $\bm w^*$ solution achieves $f_B \approx 1.5\times 10^{-7}$, using nonzero dipole magnitudes for only $14.6\%$ of the permanent magnets. It is interesting to note that $\bm w^*$ produces five times the $f_B$ value compared to $\bm m^*$. Yet the $\bm m^*$ and $\bm w^*$ distributions appear at first to be virtually identical in Fig.~\ref{fig:muse_histograms}, and in 3D visualization as in the top row of Fig.~\ref{fig:muse_summary}. Zooming into the origin of Fig.~\ref{fig:muse_histograms}, we can see that $\bm m^*$ actually breaks the grid-alignment of the dipoles in $\bm w^*$, with a large number of very weak components. Summing over the absolute values of all of these very weak but nonzero components actually amounts to the equivalent contribution of approximately 600 ($\times 4$ for symmetries) full-strength, grid-aligned permanent magnets (modulo the important directionality but this is a rough estimate anyways). This ``straggler'' issue is seen elsewhere in sparse regression, and in fact motivates the post-processing and post-optimization that is performed to improve FAMUS permanent magnet solutions. It is a substantial strength that $\bm w^*$ does not suffer from this issue, but it comes at the cost of increased $f_B$ error. 
Moreover, this sensitivity to very weak, nonzero components in the otherwise grid-aligned dipoles motivates a stochastic optimization approach, outlined in Appendix~\ref{sec:appendix_stochastic}, in order to search for configurations that are robust to small changes in the solution. 

\begin{figure*}
    \centering
    \begin{overpic}[width=0.95\linewidth]{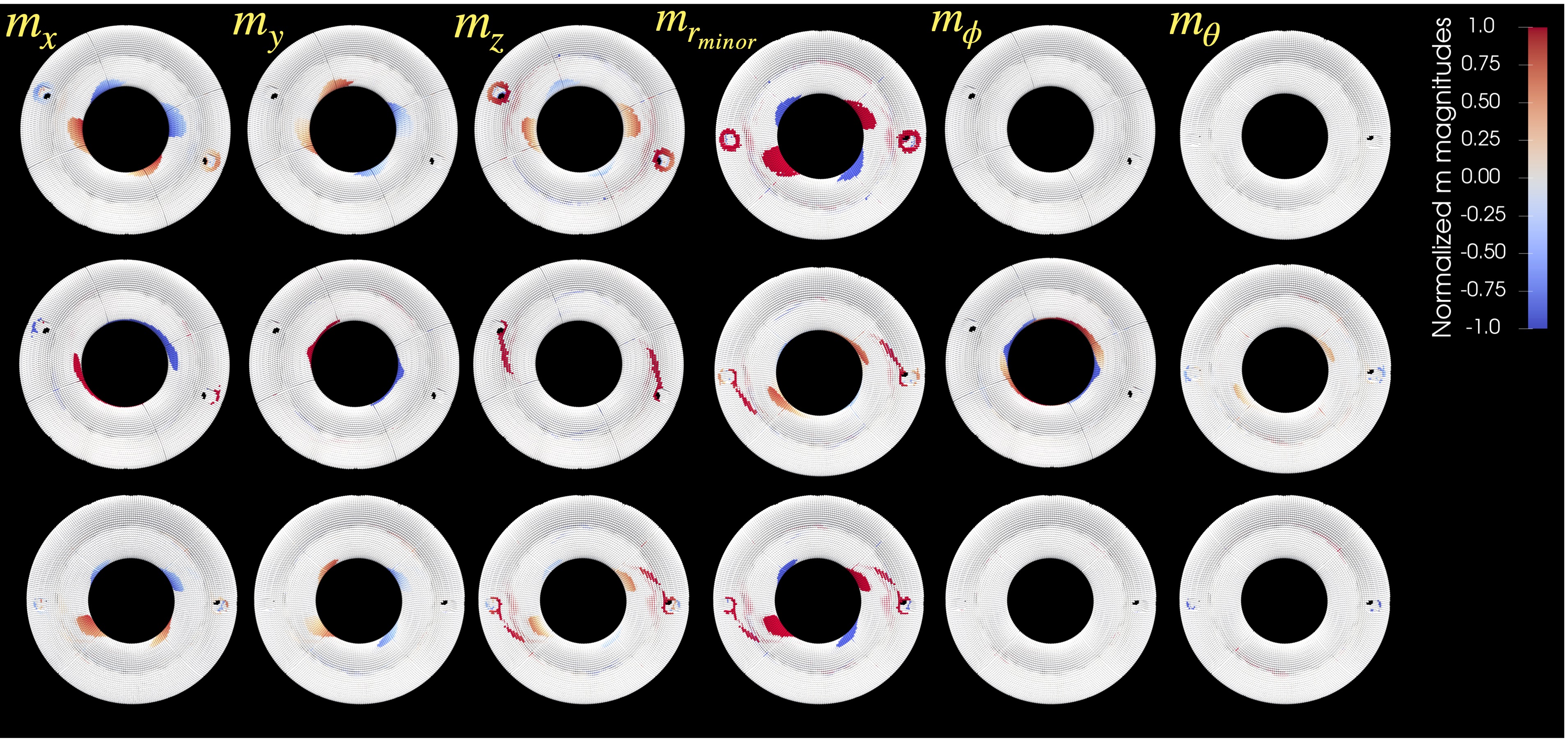}
    \put(-8.5, 34){FAMUS}
    \put(-13, 20){\parbox{1in}{ SIMSOPT \\ (Cartesian)}}
    \put(-13, 8){\parbox{1in}{ SIMSOPT \\ (toroidal)}}    
    \end{overpic}
    \caption{The different components of the MUSE $\bm m$ solutions in Cartesian and simple toroidal coordinates illustrate the different grid-alignments used in FAMUS (top row) and in SIMSOPT (middle and bottom rows are the relax-and-split dipole solutions grid-aligned with Cartesian coordinates and grid-aligned with simple toroidal coordinates, respectively). The FAMUS magnets are constrained to point inwards or outwards in the minor radial direction, and in the toroidal  relax-and-split case, magnets point in one of the three simple toroidal directions yet qualitatively reproduce the magnet structures seen by FAMUS.}
    \vspace{0.1in}
    \label{fig:grid_alignment}
\end{figure*}

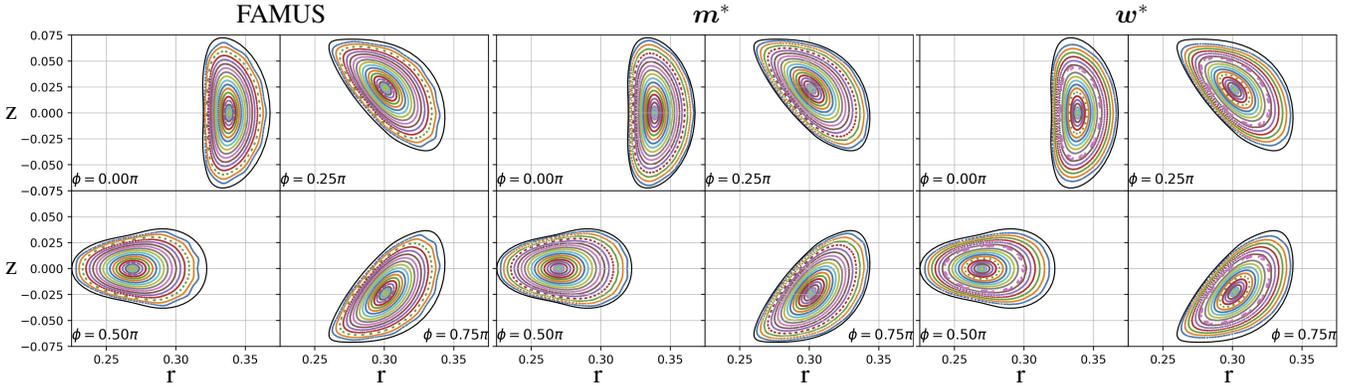
\begin{figure*}
    \centering
    \begin{overpic}[width=0.98\linewidth]{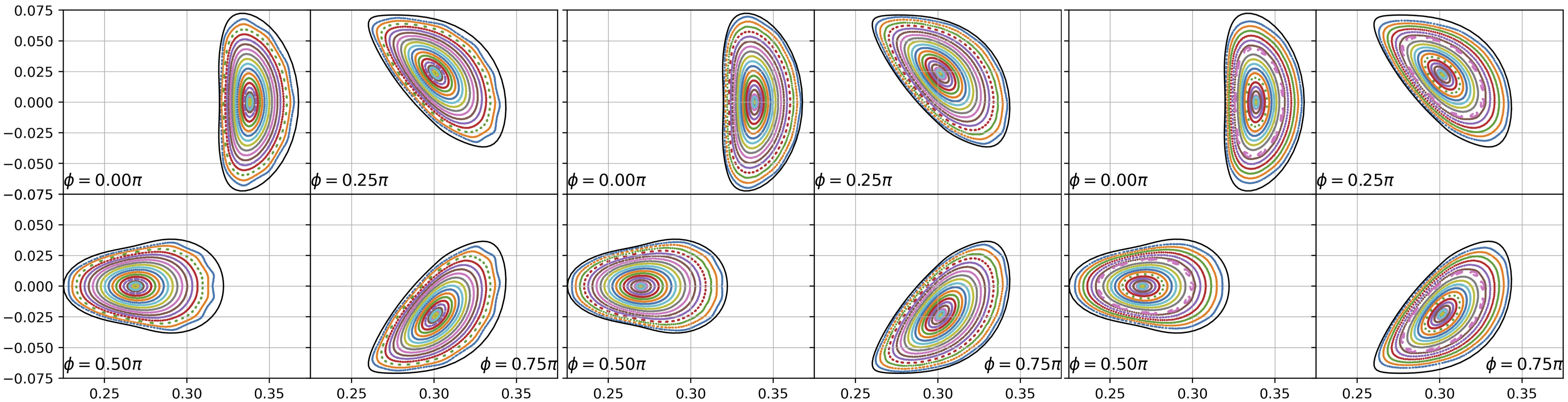}
    \put(16.5, 26){FAMUS}
    \put(51, 26){$\bm m^*$}
    \put(83, 26){$\bm w^*$}
    \put(-1, 18.75){z}
    \put(-1, 7){z}
    \put(11.2, -1.25){r}
    \put(27.1, -1.25){r}
    \put(43.5, -1.25){r}
    \put(59.2, -1.25){r}
    \put(75.5, -1.25){r}
    \put(91.25, -1.25){r}
    \end{overpic}
    \caption{Poincar\'{e} plots for MUSE solutions with the target plasma boundary in black.}
    %\vspace{-0.2in}
    \label{fig:poincare}
\end{figure*}

FAMUS uses approximately $16.8\%$ of the permanent magnets, or exactly 6608 additional magnets$^*$\blfootnote{$^*$The experimental MUSE device actually uses 9,736 magnet ``towers''. The total magnet count is much lower than is indicated by the grid of $75,460 \times 4$ locations because the final permanent magnet configuration for MUSE is generated by concatenating together identical toroidal slices (each modeled as an ideal dipole) into towers. For instance, if there are 14 slices next to each other, each 1/16" thick, a single 7/8" thick magnet is used instead. Future work could investigate promoting solutions with large, continuous groups of magnets so that this concatenation can be done effectively. 
Here, the FAMUS and relax-and-split solutions are compared by quoting the difference in the total number of permanent magnets with reference to the original grid locations. The most intuitive metric might be the effective permanent magnet volume in Eq.~\eqref{eq:eff_vol}. 
The relax-and-split solution uses $13\%$ less effective permanent magnet volume than the FAMUS solution.} 
with respect to the relax-and-split solution $\bm w^*$. 
Additional hyperparameter tuning could be done to get a comparison with FAMUS using the exact same number of dipoles, but the relax-and-split solution is already sparse, high-quality, and therefore suitable enough for this example configuration. The extra FAMUS post-processing precludes a perfect comparison with the relax-and-split method anyways. Moreover, as is illustrated in Fig.~\ref{fig:grid_alignment}, the FAMUS solution constrains the dipoles to point in the minor radial direction, while the relax-and-split solution produces dipoles that are aligned with one of the three simple toroidal directions. Despite these extra degrees of freedom, the solution qualitatively reproduces the dipole structures seen in the FAMUS solution. In contrast, the relax-and-split solution with Cartesian grid-aligned dipoles in Fig.~\ref{fig:grid_alignment} produces another excellent solution but looks qualitatively quite different. The relax-and-split and FAMUS solutions for both examples are summarized in Table~\ref{tab:results_summary} and Poincar\'{e} plots for each of the solutions are illustrated in Fig.~\ref{fig:poincare}. The surfaces generated with $\bm m^*$ look slightly improved over those of FAMUS but the $\bm w^*$ surfaces look slightly degraded.

\begin{table*}[t]
\centering
\begin{tabular}{ |c|c|c|c|c|c|  }
 \hline
 Stellarator & Solution & $f_B$ & $V_\text{eff}$ (m$^3$) & $f_{0.01}$ & Grid-alignment\\ 
  \hline
  NCSX & FAMUS & $1.6\times 10^{-6}$ & $2.29$  & 0.57 & None\\
  NCSX & SIMSOPT $\bm m^*$ & $1.6\times 10^{-6}$ & 2.34  & 0.84 & cylindrical\\
  NCSX & SIMSOPT $\bm w^*$ & $4.7\times 10^{-4}$ & 2.30  & 1 & cylindrical\\
  MUSE & FAMUS & $5.7\times 10^{-8}$ & $3.245\times 10^{-3}$  & 1 & $r_\text{minor}$\\
  MUSE & SIMSOPT $\bm m^*$ & $2.8\times 10^{-8}$ & $2.906\times 10^{-3}$ & 0.92 & toroidal\\
  MUSE & SIMSOPT $\bm w^*$ & $1.5\times 10^{-7}$ & $2.822\times 10^{-3}$  & 1 & toroidal\\
 \hline
\end{tabular}
\caption{Summary of the comparison between FAMUS and relax-and-split on the NCSX and MUSE examples.}
%\vspace{-0.2in}
\label{tab:results_summary}
\end{table*}

Consider the remarkable fact that the FAMUS and relax-and-split toroidally-aligned solutions qualitatively match. The two algorithms have different degrees of freedom and solve very different nonconvex optimization problems, yet the solutions are qualitatively very similar. There are two effects that seem like plausible reasons for this convergence between algorithms. First, certain grid locations may be essential for properly minimizing $f_B$, e.g. locations near highly-shaped magnetic fields. Second, the requirement of maximum strength, binary magnets effectively regularizes much of the permanent magnet optimization space. In other words, there may be a vast number of permanent magnet configurations that minimize $f_B$ to high-performance levels, but far fewer such configurations have binary distributions.

\section{Conclusion}\label{sec:conclusion}
We have shown that permanent magnet optimization can be formulated as sparse regression. This scientific problem is common to many fields, including automobile manufacturing, MRI, ship de-gaussing, and many other applications.
We have classified a number of loss terms that can be effectively used in permanent magnet optimization, and proposed a relax-and-split algorithm that takes advantage of the partial convexity, addresses an important class of nonsmooth and nonconvex optimization loss terms, and allows for additional convex equality and inequality constraints to be added to future objective functions. The algorithm is generally applicable, and our implementation is  computationally efficient and open-source. It can be immediately used or extended for winding-surface stellarator coil optimization 
and high-dimensional sparse regression across many scientific domains. We concluded by discovering accurate and sparse solutions for two stellarators. Interestingly, the relax-and-split algorithm qualitatively matches the FAMUS solution for the MUSE stellarator example, despite the significant differences between the algorithms and their associated optimization problems. This may be an indication that the requirement of sparse, binary magnet solutions may regularize the optimization enough to leave only a few high-quality minima, despite the nonconvexity of the problem.

Future work should further explore the parameter space of possible permanent magnet configurations on a variety of stage-1 plasma equilibria. More sophisticated physical and engineering objectives can be built into the optimization by modeling the permanent magnet deviation from a dipole field, adding constraints for the demagnetization upper bound in Eq.~\eqref{eq:coercivity_constraint}, calculating the errors induced by the permeability of the permanent magnet material, and using stochastic optimization as outlined in Appendix~\ref{sec:appendix_stochastic}. Soft rather than hard constraints on the dipole magnitudes might allow for a significantly faster convex step in the iterative relax-and-split algorithm, although we found empirically that often high-quality configurations can be computed without requiring that full convergence is achieved during each convex solve. 

Addressing the remaining optimization questions is an important task -- beyond this barrier, permanent magnet stellarators would be ideal laboratory experiments that can be constructed easily and inexpensively, while providing vital small-scale stellarator insights for informing the design of full-scale stellarators that can have construction costs in the billions of dollars. 

\section*{Acknowledgements}
AAK would like to acknowledge important conversations with Christopher Hansen, Aleksandr Aravkin, and Steven Brunton.
Thanks to Caoxiang Zhu for help with FAMUS. This work was supported by the U.S. Department of Energy under award DEFG0293ER54197 and through a grant from the Simons
Foundation (560651, ML). This research used resources of the National Energy Research Scientific Computing Center (NERSC), a U.S. Department of Energy Office of Science User Facility located at Lawrence Berkeley National Laboratory, operated under Contract No. DE-AC02-05CH11231.

\appendix
\section{Least-squares form of $f_B$}\label{sec:appendix_least_squares}
Here we derive how the $f_B$ term can be put into a least-squares form, following Zhu et al.~\cite{zhu2020topology} but with some minor changes. First, we define the geometric factor,
\medmuskip=0mu
\thinmuskip=0mu
\thickmuskip=0mu
\begin{align}
    \label{eq:geo_factor}
    \bm g_i(\phi_{i_\phi}, \theta_{i_\theta}) = \frac{\mu_0}{4\pi}\left(\frac{3\bm r_i \cdot \hat{\bm n}}{\|\bm r_i\|_2^5}\bm r_i - \frac{\hat{\bm n}}{\|\bm r_i\|_2^3} \right)\sqrt{\Delta \phi_{i_\phi} \Delta \theta_{i_\theta}\|\bm n\|_2}.
\end{align}
\medmuskip=4mu
\thinmuskip=4mu
\thickmuskip=4mu
Here $(\phi_{i_\phi}, \theta_{i_\theta})$ define a set of quadrature points on the plasma surface in toroidal coordinates, $(i_\phi, i_\theta)$ index these points, $(\Delta\phi_{i_\phi}, \Delta\theta_{i_\theta})$ denote the grid spacing, and $\bm n $ is shorthand for the normal vector at these locations, $\bm n_{i_\phi, i_\theta}$. The number of points in each direction are $N_\phi$ and $N_\theta$.
Stack all of these factors together for an overall matrix $\bm A \in \mathbb{R}^{N_\phi N_\theta\times 3D}$,
\medmuskip=0mu
\thinmuskip=0mu
\thickmuskip=0mu
\begin{align}
    \label{eq:geo_factor_stacked}
    \bm A(\phi_{i_\phi}, \theta_{i_\theta}) &= \begin{bmatrix}
    \bm g_1 & ... & \bm g_D
    \end{bmatrix} \\ \notag &= \begin{bmatrix}
    g_{1}^x & g_{1}^y & g_{1}^z & ... & g_{D}^x & g_{D}^y & g_{D}^z
    \end{bmatrix} \\ \notag &= \begin{bmatrix}
    g_{1}^x(\phi_1, \theta_1) & g_{1}^y(\phi_1, \theta_1) & g_{1}^z(\phi_1, \theta_1) & \cdots \\
    \vdots & \vdots & \vdots &  \\
    g_{1}^x(\phi_{N_\phi}, \theta_{N_\theta}) & g_{1}^y(\phi_{N_\phi}, \theta_{N_\theta}) & g_{1}^z(\phi_{N_\phi}, \theta_{N_\theta}) & \cdots \\
    \end{bmatrix}.
\end{align}
\medmuskip=4mu
\thinmuskip=4mu
\thickmuskip=4mu
This factor encodes all of the geometry, is independent of the dipoles $\bm m_i$, and can be computed once before optimization begins. To see why this is a useful definition, note that if $\bm m = [m^x_1, m^y_1, m^z_1, m^x_2, ..., m^z_D]$, we can write,
\begin{align}
\label{eq:f_b}
    f_B &= \int \left(\left(\bm B_M + \bm B_P + \bm B_C\right)\cdot\hat{\bm n}\right)^2 d^2x, \\ \notag
    &= \sum_{i_\phi}\sum_{i_\theta}\Delta\phi_{i_\phi}\Delta\theta_{i_\theta}\left(\bm B_M \cdot\hat{\bm n} - B_n\right)^2\|\bm n\|_2,
\end{align}
where the normal component of the coil plus plasma magnetic fields (on the plasma surface) is denoted $-B_n \in \mathbb{R}^{N_\phi N_\theta}$ here. Then convert the double sum to a single sum with the relabeling $i = i_\phi + N_\phi (i_\theta - 1)$, 
\begin{align}
\label{eq:f_b_simplified}
    f_B
    &= \sum_{i_\phi}\sum_{i_\theta}\left(\left(\bm B_M \cdot\hat{\bm n} - B_n\right)\sqrt{\Delta\phi_{i_\phi}\Delta\theta_{i_\theta}\|\bm n\|_2} \right)^2 \\ \notag 
    &= \sum_{i}\left(A_{ij} m_j - b_i\right)^2 = \|\bm A \bm m - \bm b\|_2^2.
\end{align}
Therefore, $f_B$ can be written as a least-squares term in the $\bm m$ vector representing the optimization variables. 
We can enforce field-period and stellarator symmetries into these equations while using only the original $\bm m$ representing a half-period of the permanent magnet manifold. The coordinate system for the permanent magnet manifold is chosen to inherit the symmetries. 
For $\bm B_M$ to exhibit field-period symmetry, $\bm m$ is rotated so that the angle between $\bm m$ and $\bm r_i$ is preserved at the new point.

If both coordinate systems are also stellarator symmetric, it follows that the new vector $\bm r_i' = [x_i, -y_i, -z_i]$ is the transformation of $\bm r_i$ under the symmetry. If each $\bm m_i$ is chosen to be stellarator symmetric, then $\bm m'_i = [-m_x, m_y, m_z]$ and this is sufficient for $\bm B_M$ to be stellarator symmetric. 

The transformation of $\bm A$ is the inverse of the transformation of $\bm m$. For instance, if the new vector $\bm m'_i$ is the result of flipping the $m_r$ component from a reflection $\bm R_S$ via stellarator symmetry, followed by a rotation $\bm R_{fp}$ by the appropriate angle for field-period symmetry, then $\bm A$ should be rotated by the negative angle, and then the resulting vector should be transformed by the stellarator symmetry. 
In total then, the least-squares term $\bm A \bm m$ that is actually used in the code is 
\begin{align}
    \left[\bm A(\bm r_i) + \bm A(\bm R_S\bm r_i)\bm R_S^T + \bm A(\bm R_{fp}\bm r_i) \bm R^T_{fp} + ...\right]\bm m,
\end{align}
which now is calculating the contributions of all of the magnets (around the full torus) to a half-period of the plasma surface, while only using the half-period variables $\bm m$.

\section{The $l_0$ proximal operator}\label{sec:appendix_prox}
Proximal operators are useful tools for optimization over nonsmooth and/or nonconvex terms that otherwise have relatively simple structure.
The proximal operator is defined as,
\begin{align}
    prox_{\alpha N}(\bm m) \equiv \argmin_{\bm y}\left[\frac{1}{2}\|\bm m - \bm y \|_2^2 + \alpha N(\bm y)\right].
\end{align}
The advantage of this formulation is that it can be analytically or rapidly numerically computed for a number of common nonconvex functions. The proximal operator of the $l_0$ ``norm'' is well-known as \textit{hard-thresholding},
\begin{align}
\label{eq:prox_l0}
    prox_{\alpha \|(.)\|_0}(\bm m) &= \mathcal{H}_{\sqrt{2\alpha}}(m_1)\times ... \times \mathcal{H}_{\sqrt{2\alpha}}(m_{3D}), \\
\label{eq:hard_threshold_operator}
    \mathcal{H}_{\sqrt{2\alpha}}(m_i) &= \begin{cases}
    0 & |m_i| < \sqrt{2\alpha} \\
    m_i & |m_i| \geq \sqrt{2\alpha}
    \end{cases}.
\end{align}
In practice, we actually hard threshold the normalized $\bm m$ (note that the $l_0$ norm is invariant to a global rescaling of all the $\bm m$ components before optimization begins), since otherwise the hard thresholding will preferentially remove small magnets, even if those magnets are at full strength and very important to the solution. The maximum strength of the permanent magnets is usually proportional in some way to the geometry (for instance, directly proportional to the cylindrical radius), and therefore an unnormalized hard-thresholding would tend to remove the geometrically-small magnets first. 

\section{Stochastic optimization with permanent magnets}\label{sec:appendix_stochastic}
It may not be immediately clear that the relax-and-split algorithm can be extended for more advanced optimization techniques such as stochastic optimization. Stochastic optimization is a useful technique for generating configurations that are robust to random and systematic errors in the coil shapes (traditional coil optimization) and the permanent magnet magnitudes and directions (permanent magnet optimization). In this section, we briefly illustrate that typical forms of stochastic optimization preserve all the loss term structure needed for relax-and-split. Consider a vector of independent, normally-distributed random variables $\bm \xi$ as in Wechsung et al.~\cite{wechsung2021single}. The goal is to minimize the expectation $\mathbb{E}$ of the $\bm B\cdot\hat{\bm n}$ errors over the distribution of perturbations $\bm \xi$,
\begin{align}
\label{eq:sopt1}
    \min_{\bm m} \mathbb{E}(f_B(\bm m + \bm \xi)) + \lambda R(\bm m) + \alpha N(\bm m).
\end{align}
Alternatively, one can optimize for the worst case scenario,
\begin{align}
\label{eq:sopt2}
    \min_{\bm m}\max_{\bm \xi} \mathbb{E}(f_B(\bm m + \bm \xi)) + \lambda R(\bm m) + \alpha N(\bm m).
\end{align}
The point of the regularization terms is to simplify the manufacturing, so they are relevant \textit{before} any perturbations are introduced. For instance, if a dipole has zero magnitude, a  magnet will not be placed, and therefore there are no manufacturing errors to consider at this location.
The expectation value can be approximated using the sample average approximation; drawing $S$ independent realizations of $\bm \xi$ and using,
\begin{align}
\label{eq:expectation_approximation}
    \mathbb{E}(f_B(\bm m, \bm \xi)) \approx \frac{1}{S}\sum_{s=1}^S f_B(\bm m + \bm \xi_s),
\end{align}
to achieve $\mathcal{O}(S^{-\frac{1}{2}})$ approximation error. The $\bm \xi_s$ are fixed, so~\eqref{eq:sopt1} and~\eqref{eq:sopt2} are optimizations using deterministic sums of the original objectives. Sums of convex terms are also convex, and thus relax-and-split can be applied in this setting. 

 \bibliography{PM}

\end{document}